# Estimating individual employment status using mobile phone network data

[1]Pål Sundsøy, [1]Johannes Bjelland, [1]Bjørn-Atle Reme, [2]Eaman Jahani, [3,4]Erik Wetter, [3]Linus Bengtsson

[1]Telenor Group Research, [2]MIT Media Lab, [3]Flowminder Foundation, [4]Stockholm School of Economics

## ABSTRACT

This study provides the first confirmation that individual employment status can be predicted from standard mobile phone network logs externally validated with household survey data. Individual welfare and households' vulnerability to shocks are intimately connected to employment status and professions of household breadwinners. At a societal level unemployment is an important indicator of the performance of an economy. By deriving a broad set of novel mobile phone network indicators reflecting users' financial, social and mobility patterns we show how machine learning models can be used to predict 18 categories of profession in a South-Asian developing country. The model predicts individual unemployment status with 70.4% accuracy. We further show how unemployment can be aggregated from individual level and mapped geographically at cell tower resolution, providing a promising approach to map labor market economic indicators, and the distribution of economic productivity and vulnerability between censuses, especially in heterogeneous urban areas. The method also provides a promising approach to support data collection on vulnerable populations, which are frequently under-represented in official surveys.

## Keywords
Big-Data Development, machine learning, unemployment, socio-economic indicators, mobile phone metadata, profession

## 1. INTRODUCTION

Unemployment is a key indicator of labor market performance [1,2]. When workers are unemployed, their families also get affected, while the nation as a whole loses their contribution to the economy in terms of the goods and services that could have been produced [3]. Unemployed workers also lose their purchasing power, which can lead to the unemployment for other workers, creating a cascading effect that ripples through the economy [4]. Additionally, unemployment has been shown to be a driver of interregional migration patterns [5].

Counting each and every unemployed person on a monthly basis would be a very expensive, time-consuming and impractical exercise. In many countries, such as US, a monthly population survey is run to measure the extent of unemployment in the nation [6]. In developing countries such surveys often tend to have a low spatial and temporal frequency [7]. Lacking statistics may lead to higher uncertainties in economic outlook, lower purchasing capacity and higher burden of debt. The problems of unemployment and poverty have always been major obstacles to economic development [8], and proper background statistics is important to change this trend.

The increasing availability and reliability of new data sources, and the growing demand of comprehensive, up-to-date international employment data are therefore of high priority. Specifically, privately held data sources have been shown to hold great promise and opportunity for economic research, due to both high spatial and temporal granularity [9].

One of the most promising rich new data sources is mobile phone network logs [10], which have the potential to deliver near real-time information of human behavior on individual and societal scale [11]. The prediction from mobile phone metadata are vast given that more than half of the world's population now own a mobile phone. Several research studies have used large-scale mobile phone metadata, in the form of call detail records (CDR) and airtime purchases (top-up) to quantify various socio-economic dimensions. On aggregated level mobile phone data have shown to provide proxy indicators for assessing regional poverty levels [12,13], illiteracy [14], population estimates [15], human migration [16,17] and epidemic spreading [18]. On individual level mobile phone data have been used to predict, among others, socio-economic status [19,20], demographics [21,22] and personality [23].

Two previous papers analyze employment trends through cell phone data. The work by [24] argue that unemployment rates may be predicted two-to-eight weeks prior to the release of traditional estimates and predict future rates up to four months ahead of official reports accurately than using historical data alone. The other study [25] shows that mobile phone indicators are associated with unemployment and this relationship is robust when controlling for district area, population and mobile penetration rate. The results of these analyses highlight the importance of investigating the relationship between mobile phone data and employment data further.

Our work separates from [24,25] in several ways:

(1) **Bottom-up approach:** we focus on predicting employment status on the *individual* level to be able get a clean view of the main drivers and ability to discover global predictors that are useful across various employment groups. This is of relevance also as previous research has uncovered non-linear relationships between worker flows and job flows at the micro level, indicating a more complex relationship between the micro and macro levels of employment statistics than simple aggregation [26].
(2) **Geography:** We focus on a low HDI South-Asian developing country where employment statistics is almost non-existing and highly in demand.
(3) **Method:** Individually matched mobile phone data and large-scale survey data allow us to carefully optimize machine-learning models for multiple professions.
(4) **Data:** In addition to CDR data we also include airtime purchases, as financial proxy, for our analysis.

The rest of this paper is organized as follows: In section 2 we describe the methodological approach, including the features and modeling approach. In section 3 we describe the results. In chapter 4 we discuss the limitations from a holistic perspective, while we finally draw our conclusions.

## 2. APPROACH

### 2.1 Data

**Household survey data:** Data from two nationally representative cross-sectional household surveys of 200,000 individuals in a low-income South Asian country was analyzed. The data was collected for business intelligence purposes at time Q114 and Q214 by an external survey company commissioned by the operator. The survey discriminated between 18 types of professions for the head of household, including currently being retired and unemployed. The head of household's was asked for his or her most frequently used phone number. 87% of households in the country has at least one mobile phone.

**Mobile phone data:** Mobile phone logs for 76 000 of the surveyed 200 000 individuals belonging to the leading operator were retrieved from a period of six months and de-identified by the operator. Individual level features were built from the raw mobile phone data and was subsequently coupled with the corresponding de-identified phone numbers from the survey. The social features were subsetted from a graph consisting of in total 113 million subscribers and 2.7 billion social ties. No content of messages or calls were accessible and all individual level data remained with the operator.

The following sub-sections describe the features and the machine learning algorithm used for our prediction.

### 2.2 Features

**Independent variables:** The independent variables are built entirely from the cell phone datasets. A structured dataset consisting of 160 novel mobile phone features are built from the raw CDRs and airtime purchases, and categorized into three dimensions: (1) financial (2) mobility and (3) social features (Table 1). The features are customized to be predictive of employment status and include various parameters of the corresponding distributions such as weekly or monthly median, mean and variance. In addition to basic features such as incoming and outgoing MMS, voice, SMS, internet and video calls we investigate more customized features such as the consumption rate of airtime purchases (spending speed), the amount time spent on each base station, the size of social circle, the time spent on different contacts and features related to the phone type owned by the customer.
**Dependent variables:** The dependent variables were built from the survey data. Since our aim is to separate one specific profession/employment status from the others we build 18 binary classifiers – one for each profession (student/non-student, employed/unemployed and so forth). These classifiers are then trained separately.

**Table 1. Sample of independent features from mobile phone metadata used in model**

| Dimension | Features |
| --- | --- |
| Financial | **Airtime purchases:** Recharge amount per transaction, Spending speed, fraction of lowest/highest recharge amount, coefficient of variation recharge amount etc<br>**Revenue:** Charge of outgoing/incoming SMS, MMS, voice, video, value added sevices, roaming, internet etc.<br>**Handset:** Manufacturer, brand, camera enabled, smart/feature/basic phone etc |
| Mobility | Home district/tower, radius of gyration, entropy of places, number of places visited etc. |
| Social | **Social Network:** Interaction per contact, degree, entropy of contacts etc.<br>**General phone usage:** Out/In voice duration, SMS count, Internet volume/count, MMS count, video count/duration, value-added services duration/count etc. |

### 2.3 Deep Learning Models

We tested our features against several algorithms such as gradient boosted machines (GBM), random forest (RF), support vector machines (SVM) and K-nearest neighbors (kNN). Based on the performance of individual algorithms we propose a standard multi-layer feedforward neural network architecture where the weighted combination of the n input signals is aggregated, and an output signal $f(\alpha)$ is transmitted by the connected neuron. The function f used for the nonlinear activation is rectifier $f(\alpha) \approx \log(1+e^{\alpha})$. To minimize the loss function we apply a standard stochastic gradient descent with the gradient computed via back-propagation. We use dropout as a regularization technique to prevent over-fitting [27]. Dropout secures that each training example is used in a different model which all share the same global parameters. For the input layer we use the value of 0.1 and 0.2 for the hidden layers. In total 18 models are built – one for each pre-classified profession type.

To compensate class imbalance, the minority class in the training set is up-sampled. The minority class is then randomly sampled, with replacement, to be of the same size as the majority class. In our set-up, each model is trained and tested using a 75%/25% (training/testing) split. Commonly used performance metrics for classification problems [28], including overall accuracy, sensitivity and specificity, are reported for the test-set.

## 3. RESULTS

### 3.1 Individual employment status

The average prediction accuracy for all 18 profession groups were 67.5%, with clerk being the easiest to predict with accuracy of 73.5%, and skilled worker the most difficult (accuracy: 61.9%) (Fig. 1a).

Our unemployment model predicted whether phone users were unemployed with an accuracy of 70.4% (95% CI:70.1-70.6%). The accuracy difference between the training and test-set was 3.6%, which indicate our trained model has good generalization power. The true positive rate (sensitivity/recall) was 67% and true negative rate (specificity) 70.4%. Given the original baseline of 2.1%, the model predicts unemployment on average 30 times better than random.

Each cross-validated model was subsequently restricted to use its 20 most important predictors (applying Gedeon method [29]). An investigation of the five most important predictors for each profession is given in Fig 2b. This network shows how the professions are linked together via common predictors. We observe that several features are predictive across multiple professions – indicated by high in-degree in the network. Predictors that are superior across multiple professions include the most frequently used cell tower (longitude and latitude): in the case of unemployed (red node) this signal indicate that the model may catch regions of low economic development status, e.g. slum areas where unemployment is high. Other cross-profession predictors include number of visited places, the radius of gyration (how far the person usually travels from his home tower) and recharge amount per transaction. These indicators have earlier been shown to be important financial proxy indicators for household income in underdeveloped Asian markets [20]. Unemployed people tend further to have few interactions with their friends, generate more voice calls at night (when calls are cheaper) and make less voice calls. They also tend to top-up with the lowest recharge amount per transaction (sixth predictor) - a feature that also occurs as a predictor of low household income [20].

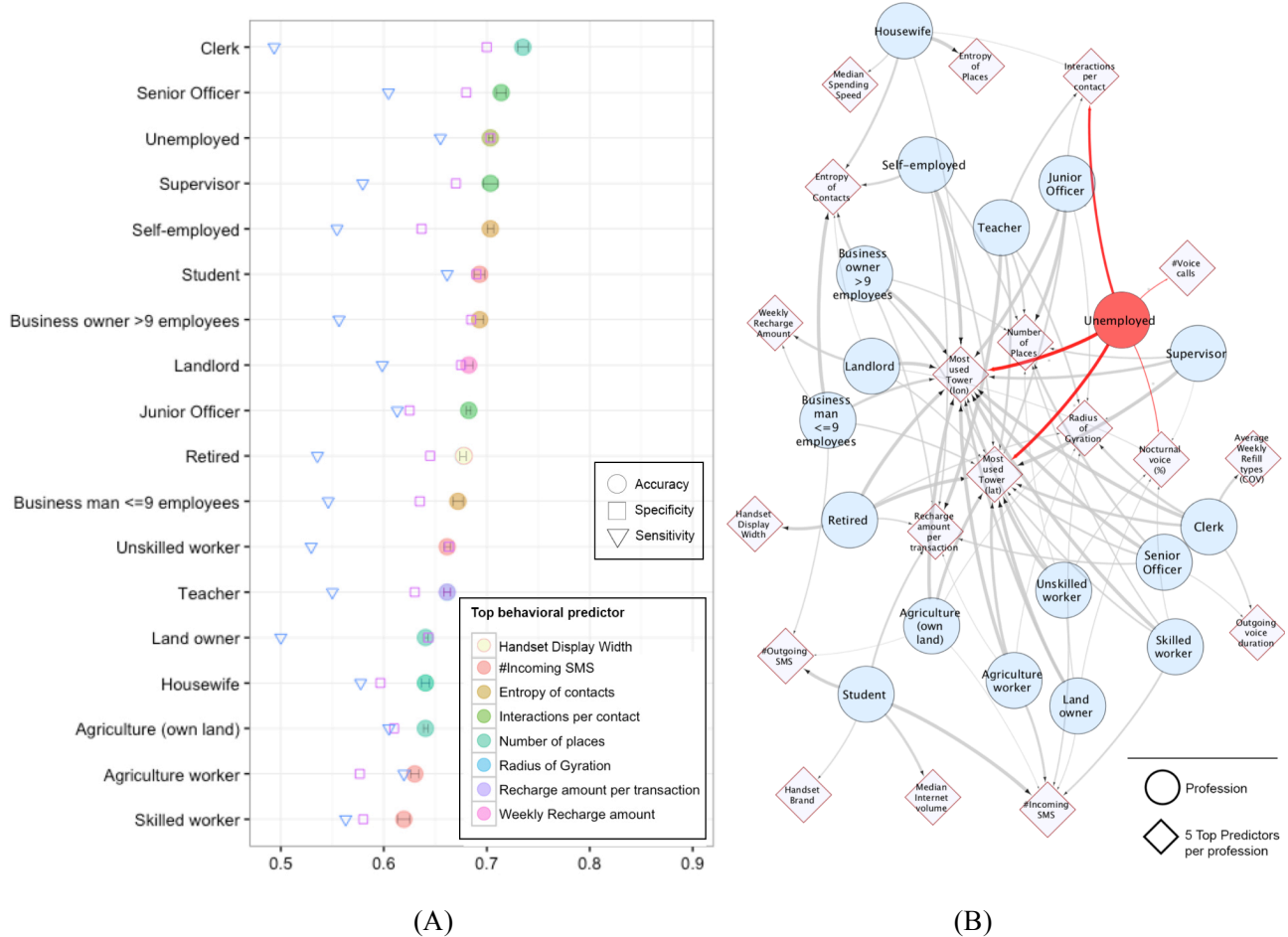

(A)          (B)

Figure 1: A) **Test-set performance of predicting individual profession from mobile phone data.** For each profession accuracy ●, sensitivity ▽ and specificity □ are reported. Top behavioral predictors are indicated with colors (excluding most used tower (lon,lat)) B) **Network of professions** linked via common predictors. Each profession links to their 5 most important predictors. The link width is proportional to the scaled relative importance of the current predictor. For example the most important predictors for being unemployed (highlighted in red node color) are most used cell tower (lon,lat), interaction per contact, nocturnal voice (%) and number of voice calls.

As seen in the figure, students have the most unique predictive signal, when it comes to few overlapping predictors with other groups. Interestingly they are not the easiest group to predict, which indicate the Deep Learning models have found stronger non-linear relationships in other categories of profession.

## 3.2 Geographical employment mapping

A natural next step is to move from individual employment to geographical distribution of employment rates and professions. Large Asian cities are typically covered by thousands of mobile phone towers opening up to the possibility of providing a detailed spatial understanding of differences in employment rates and profession types. In figure 2 we have mapped out the predicted geographical distribution of employment rate per home tower, in one of the large cities for six different groups. The individual employment rates are here calculated using the test set, aggregated and averaged to the tower from which most calls were made between 7pm and 5am (defined as home tower). Fig 2a shows that the unemployed population are spatially spread out across the city, indicating many pockets of unemployment that traditional surveys would not easily pick up. The retired population (Fig 2d) and landlords (Fig 2c) are spatially more concentrated.

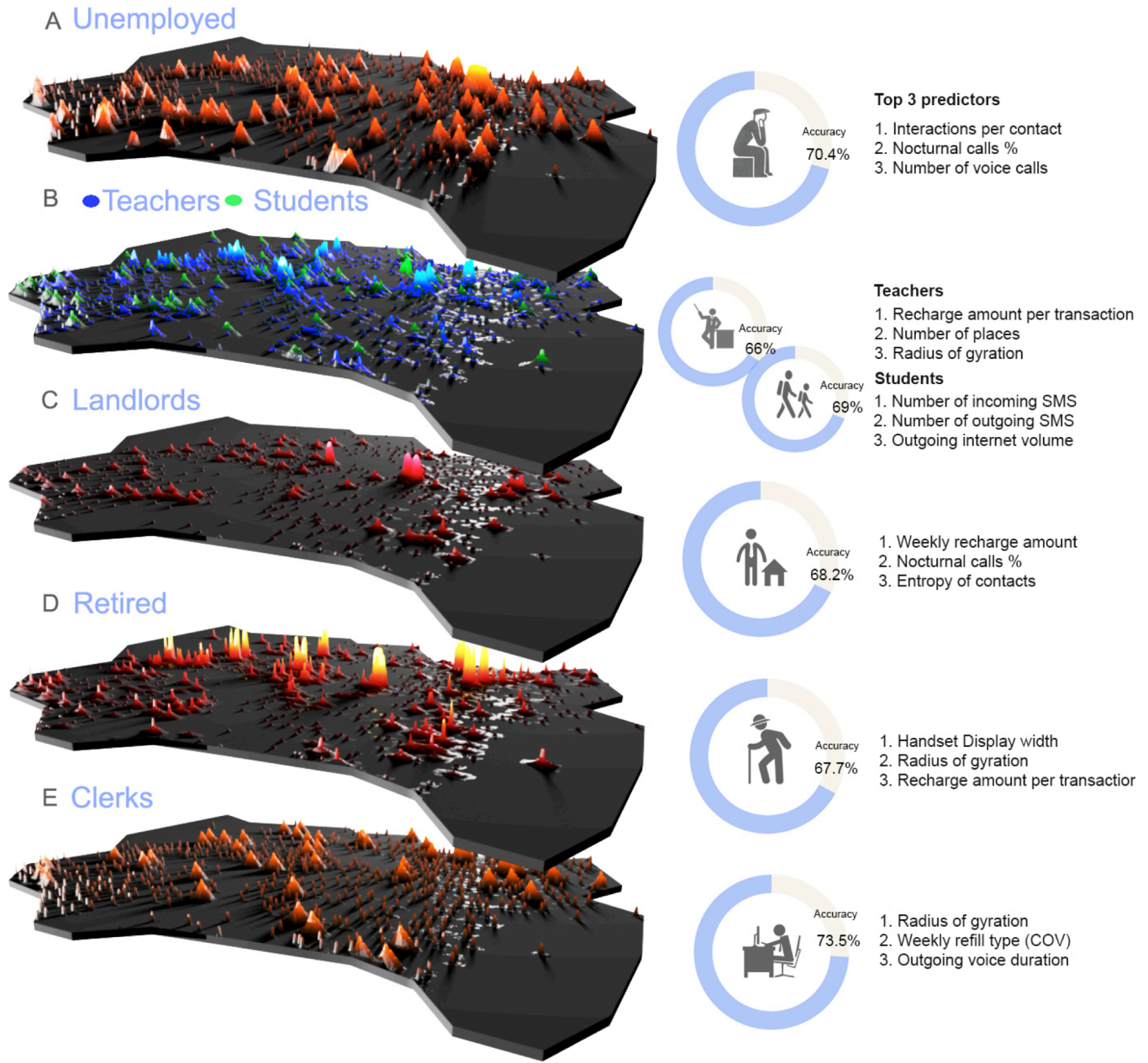

**Figure 2 Geographical distributions of employment status and profession categories** per base station in one of the larger Asian cities with over 1,500 cell towers and 18 million people. Employment rates are calculated by using the out-of-sample test set, aggregated and averaged to their respective home tower. Individual prediction accuracy and top three predictors are given to the right for A) Unemployed B) Teachers and students C) Landlords D) Retired E) Clerks.

By investigating the top 3 most important behavioral predictors we interestingly observe that the physical width of handset is the most important for the retired people. The elderly in this market tends to have older and smaller phones than the younger generation. This in contrast to Scandinavian markets where larger handsets are marketed especially towards elderly people. We also notice that SMS and mobile Internet consumption are the best indicators for being a student (Fig 2b). The most important predictor of Clerks (Fig 2e) is a low radius of gyration (or mobility radius – reflecting static office jobs). They also tend to have more advanced consumption patterns (more variation in top-up refill types) and longer voice duration. We also notice that airtime purchase information is among top predictors for several professions, stressing the importance to include such datasets in future research.

## 4. DISCUSSION

This study shows how employment status can be predicted from mobile phone logs, purely by investigating users' metadata. By deriving economic, social and mobility features for each mobile user we predict individual employment status with up to 73.5% accuracy. We address how various profession groups relates in a network via industry standard mobile network indicators, and further show how individual employment can be aggregated and mapped geographically with high spatial resolution on cell tower level. The geospatial indicators available in mobile network data additionally allow promising avenues for research on two major topics related to the economic impacts of employment status; the increased productivity effects that have been found to be an outcome of increased spatial density of employment [30], and the interregional migration patterns that are related to unemployment [5].

One general concern in such studies is always the sampling selection bias. A large data set may make the sampling rate irrelevant, but it doesn't necessarily make it representative. The fact that the people who use mobile phone are not necessarily a representative sample of the larger population considered. This issue is especially of high relevance when considering how mobile phone data may be used for monitoring, economic forecasting and development. Research studies are often based purely on data from one mobile operator, and depending on the type of data one can expect individuals to be represented disproportionally with respect to certain characteristics. These problems persist even if data from all operators in a country were available, nearing the total population. In our study we consider data from one large operator, where customers have to own a mobile phone to be counted. We argue, however, that there should be a high correlation between population employment and sample employment, since our sample is large, and most people in the country own a mobile phone (more than 85% penetration rate). Additional sources for external validation were not easily obtainable.

An important policy application of this work is the prediction of regional and individual employment rates in developing countries where official statistics is limited or non-existing.